\documentclass[a4paper]{IEEEtran}
\usepackage{ifpdf}

\ifCLASSINFOpdf
   \usepackage[pdftex]{graphicx}
   \DeclareGraphicsExtensions{.pdf,.jpeg,.png,.jpg}
\else
  \usepackage[dvips]{graphicx}
  \DeclareGraphicsExtensions{.eps}
\fi
\usepackage[cmex10]{amsmath}

\usepackage{stfloats}
\usepackage{multicol}
\usepackage{multirow}
\usepackage{amssymb}

\begin{document}
\pagestyle{empty}

%
\title{Empirically Characterizing the Buffer Behaviour of Real Devices}

\author{\IEEEauthorblockN{Luis Sequeira, Juli\'{a}n Fern\'{a}ndez-Navajas, Jose Saldana, Luis Casadesus}
\IEEEauthorblockA{Communications Technology Group (GTC)-Arag\'{o}n Inst. of Engeneering Research (I3A)\\
Dpt. IEC. Ada Byron Building. CPS Univ. Zaragoza\\
50018 Zaragoza, Spain\\
Email: \{sequeira, navajas, jsaldana, luis.casadesus\}@unizar.es}}

\maketitle
\thispagestyle{empty}
\begin{abstract}
All the routers include a buffer in order to enqueue packets waiting to be transmitted. The behaviour of the routers' buffer is of primary importance when studying network traffic, since it may modify some characteristics, as delay or jitter, and may also drop packets. As a consequence, the characterization of this buffer is interesting, especially when real-time flows are being transmitted: if the buffer characteristics are known, then different techniques can be used so as to adapt the traffic: multiplexing a number of small packets into a big one, fragmentation, etc. This work presents a preliminary study of how to determine the technical and functional characteristics of the buffer of a certain device (as e.g. behaviour, size, limits, input and output rate), or even in a remote Internet network node. Two different methodologies are considered, and tested on two real scenarios which have been implemented; real measurements permit the estimation of the buffer size, and the input and output rates, when there is physical or remote access to the ``System Under Test". In case of having physical access, the maximum number of packets in the queue can be determined by counting. In contrast, if the node is remote, its buffer size has to be estimated. We have obtained accurate results in wired and wireless networks.
\end{abstract}

\begin{IEEEkeywords}
Buffer size, queueing, unattended measurements.
\end{IEEEkeywords}

%
\IEEEpeerreviewmaketitle

\section{Introduction}

The large increase in the number of users and new multimedia services over Internet generate a significant amount of network traffic. Moreover, the expectation of future growth for the use of multimedia applications such as, e.g. Voice over Internet Protocol (VoIP), indicates that this tendency will increase. In parallel, the heterogeneous characteristics of the different Internet accesses, together with user demands, make it necessary to define the Quality of Service (QoS) that they offer, especially when the accesses provide support to real-time applications. 

Packet size may vary between different Internet services and applications: while some of them, as e.g. VoIP, generate small packet, in order of a few of tens of bytes, others use large packets in the order to improve header efficiency. In addition, traffic behaviour has significant impact on network resources: while some services inject traffic with a constant bit rate in order to provide certain quality of service and a better user's experience, other applications generate bursty traffic, with a different number of frames into each burst.

The demand of high amounts of data by Internet users, in conjunction with the existing complex network architectures, implies that some network points become critical bottlenecks. Nowadays, this mainly happens in access networks, because capabilities are lower than the ones available in the backbone; in addition, bottlenecks may also appear at critical points of high-performance networks.

In these points, the discarding of packets in router queues is the main cause of packet loss. So the implementation of router buffers, and the implemented policies, are of primary importance in order to ensure the correct delivery of the traffic of different applications and services.

Mid and low-end routers can be found in access networks, and they normally do not implement advanced traffic management techniques. However, they always use buffers as a traffic regulator mechanism so buffer size becomes an important design parameter; buffer can be measured in different ways: maximum number of packets it can store, amount of bytes, or even queueing time limit. Most Internet routers use FIFO drop-tail buffers \cite{buffers11} but there exist another techniques to manage drop-tail buffers, e.g. Random Early Detection (RED). This techniques, in conjunction with buffer size, mainly define the buffer behaviour, and therefore, how traffic is affected by it.

As a consequence, if the size of the buffer and its behaviour are known, some techniques can be used so as to improve link utilization: multiplexing a number of small packets into a big one, fragmentation, etc. But there is a problem: manufacturers do not include all the implementations details in  the technical specifications of the devices, but just part of them, related to technology used, etc. In the other hand, if a communication has to cross another network or the Internet, some knowledge about the device's characteristics or the buffer's behaviour will be interesting. For these situations, our group is currently working on the development of a tool able to discovery some characteristics of the buffer and its behaviour. The idea is to permit measurements not only when we have physical access to ``System Under Test" but also in the case of only having remote access.

The paper is organized as follows: section II discusses the
related works. The test methodology is presented in section III.
The next section covers the experimental results. The paper ends
with the conclusions.

%
\section{Related works}

\subsection{Buffer: types and behaviour}

The fact of having rates at the input and output link of routers produce bottlenecks in the network, so packet loss may occur. Buffers are used to reduce packet loss by absorbing transient bursts of traffic when routers cannot forward them at that moment. They are instrumental in keeping output links fully utilised during congestion times.

For many years, researchers accepted the so-called \textit{rule of thumb} to obtain the amount of buffering needed at a router's output interface. This rule was proposed in $ 1994 $ \cite{buffers6} and it is given by $ B=C \times RTT $, where $ B $ is the buffer size, $ RTT $ is the average round-trip time and $ C $ the capacity of the router's network interface. This \textit{rule of thumb} is also called the Bandwidth Delay Product (BDP). It was experimentally obtained using at most 8 TCP flows on a $ 40 \; Mbps $ core link, so there is no recommendation for sizing buffers when there is a significant number of TCP flows with different $ RTTs $.
 
In $ 2004 $, researchers from \textit{Stanford University} \cite{buffers7} proposed a reduced buffer size by dividing BDP by the square root of the number of TCP flows $ B=C \times RTT / \sqrt{N} $. This new approximation assumes that the number of TCP flows is large enough so as to consider them as asynchronous and independent from each other. This model was called \textit{small buffer}.

In \cite{buffers10} it was suggested the use of even smaller buffers, called \textit{tiny buffers}, considering a buffer size of some tens of packets. However, the use of this model presents a trade-off: on one hand, reducing buffers to only a few dozen of $ KBytes $ can produce a $ 10\%-20\% $ drop probability. The model was obtained based in no bursty traffic and TCP flows were not synchronised.

However, some real-time IP flows are bursty as e.g. video streaming, so this leaves some uncertainty in buffer sizing. In \cite{buffers9} and \cite{buffers5}, TCP and UDP combined traffic in very small buffers was obtained using non-bursty traffic, and finding an anomalous region for UDP packets, where loss probability grows when buffer size increase.

It has also been observed in the literature that the buffer size is measured in different ways: e.g., in \cite{buffers1} the routers of two manufactures are compared, and one gives the information in packets, whereas the other one measures it in milliseconds, which is equivalent to bytes.

In the other hand, router buffer size is not a parameter appearing in the technical specifications that manufacturers provide with their devices. However, this design characteristic is important when planning a network. The reason for this is that there is a relationship between router buffer size and link utilization, since an excessive amount of memory would generate a significant latency increment when the buffer is full. On the other hand, a very small amount of memory in the buffer will increase packet loss in congestion time. As a consequence, the knowledge of the buffer behaviour is an interesting parameter which can be considered when trying to improve link utilization. \nocite{buffers2} \nocite{buffers3} \nocite{buffers4} \nocite{buffers8} \nocite{buffers11}

\subsection{Influence of the buffer in different services}

In \cite{gtc14} the authors present a Matlab simulation study of the influence of a multiplexing method on the parameters that define the subjective quality of online games: mainly delay, jitter and packet loss. The work considers two buffer implementations, each one with two buffer sizes, in order to study the relationship between router's buffer and multiplexing on subjective quality. The multiplexed traffic of a number of users shared the same access link. The results show that small buffers present better characteristics in order to maintain delay and jitter in adequate levels, at the cost of increasing packet loss. In addition, buffers whose size is measured in packets also increase packets loss.

Another case of multiplexing study is presented in \cite{gtc15}, where three different router's buffer policies (dedicated, big and time-limited buffer) are tested, also using two multiplexing schemes. Router's buffer policies cause different behaviour in packet loss, as well as R-factor.

\section{Test methodology}

\subsection{Test procedure}

\begin{figure}
	\centering
	\includegraphics[width=3.5in]{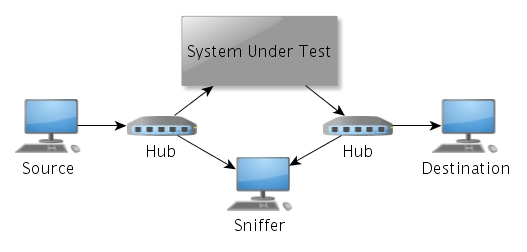}
	\caption{Topology used for test.}
	\label{fig:buffer_size_topology}
\end{figure}

The scheme of the tests is shown in figure \ref{fig:buffer_size_topology}. There is a ``System Under Test" (SUT from now), which may be either a device or a network. Traffic is sent from a source, and arrives to the destination traversing the SUT. Two hubs and a server are used in order to capture the traffic at the input and at the output of the SUT. The test is based on the sending of a burst of UDP packets from the source to the destination machine, so as to produce a buffer overflow in the SUT. This test is repeated using different bandwidths. Packets of different sizes are used so as to determine if the buffer is measured in number of packets or in bytes.
 
\subsection{Methodology}

We will use two methodologies to estimate buffer behaviour, size, limits and output rate. The methods are based on the premise that output rate can be obtained from \textit{destination} capture. Output rate depends on the technology used in each case (LAN, WiFi). 

\begin{itemize}
	\item \textbf{Method 1:} Counting the number of packets in the queue in the moment that a packet arrives at the buffer. 
	\item \textbf{Method 2:} If the delay of a packet in the buffer can be determined, then the variations of this delay can give  us useful information for estimating buffer size.
\end{itemize}

The first option brings a more accurate estimation, but it requires physical access to the SUT. The second option can also be used when there is not direct access to the system.

\subsubsection{Finding buffer size with physical access}

\begin{figure}
	\centering
	\includegraphics[width=3.5in]{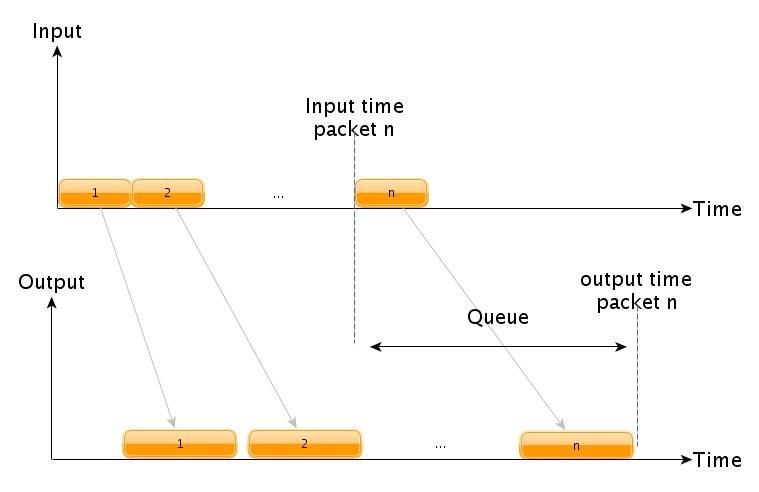}
	\caption{Estimating packets in queuing.}
	\label{fig:buffer_size}
\end{figure}

The first option, which can be used to determinate buffer size in case of having physical access is shown in figure \ref{fig:buffer_size}. All packets transmitted are identified by a sequence number included in the payload, so buffer size is estimated by the number of packets in queue between the arrival and leave time.

When physical access to the SUT is guaranteed, a \textit{sniffer} can be included, capturing traffic at the ends of the device. When the capture ends, two files are saved at the input and output of the \textit{sniffer}: we will denote them as \textit{in-capture} and \textit{out-capture}.
 
Both files are processed with a \textit{shell script} to calculate packet delay, packet loss, interarrival packet time, input and output buffer rate and filling buffer rate. Buffer size is determined for each packet as follows: for all packets in \textit{out-capture}, a \textit{shell script} looks for the incoming time in \textit{in-capture} and counts in \textit{out-capture} the number of packets between incoming time and the time stamp registered in \textit{out-capture}, finally the buffer size is estimated as the average of all these values.

\subsubsection{Finding buffer size with remote access}

A good estimation can also be obtained even if there is no physical access. By the use of ETG (E2E Traffic Generator), a tool which we will explain in the next section, unattended measurements in \textit{destination} have been deployed. In spite of not having direct \textit{sniffer} captures, it is possible to know the incoming traffic characteristics in the \textit{destination}. Nevertheless, some modifications are required in the way calculations are done. In  this case, we have used the second option to determinate buffer size.

\begin{figure*}[!t]
	\centering
	\includegraphics[width=7in]{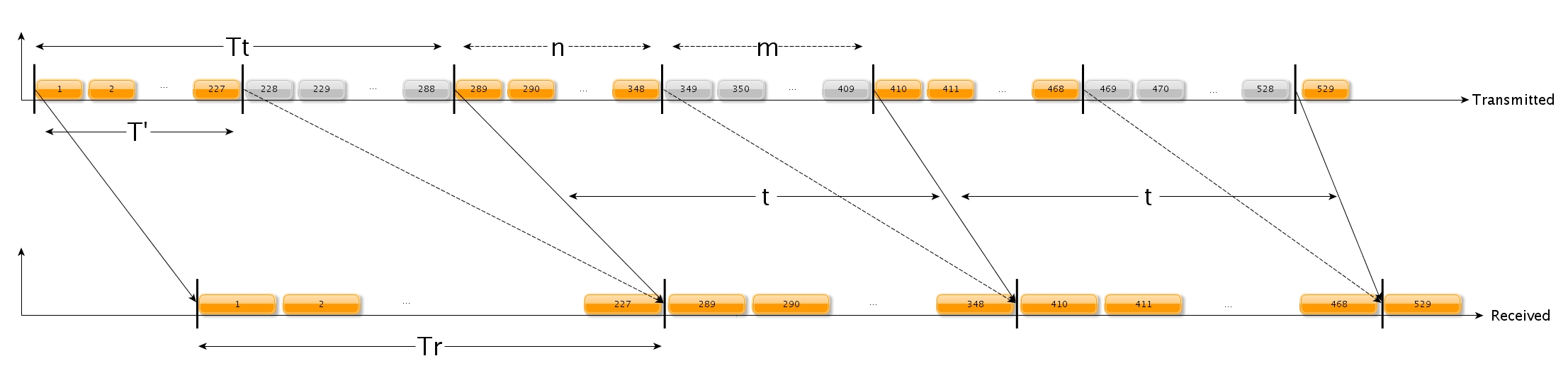}
	\caption{Estimating packets in queue in remote access.}
	\label{fig:dropping}
\end{figure*}

Figure \ref{fig:dropping} shows the relationship between sent and received packet times. The method is based on sending a continuous packet flow at a rate bigger than the output capacity. This will flood the buffer and cause packet loss. $ T' $ is the time required to completely fill the buffer at the same time it is being emptied at an output rate, and $ T_{r} $ is the sum of two times: the delay for completely filling the buffer until a packet loss occurs; plus the time the last accepted packet needs to go through the buffer. As a consequence, this time will be noticed at the out-capture when the first packet is missing. So, 

\begin{equation}
	T_{r}=T_{fill}+T_{empty}
	\label{eq1}
\end{equation}

Let $ R_{in} $ and $ R_{out} $ the input and output rates of the buffer, respectivelly. We define $ R_{fill} $ as the rate in which the buffer fills when $ R_{in} $ is bigger than $ R_{out} $ ($ R_{fill}=R_{in}-R_{out} $). $ L_{buffer} $ is the size of the buffer in bytes. A packet spends $ L_{buffer}/rate $ to cross the full buffer, so we can obtain $ T_{r} $ as,

\begin{equation}
	T_{r}=\frac{L_{buffer}}{R_{fill}}+\frac{L_{buffer}}{R_{out}}
	\label{eq2}
\end{equation}

therefore,

\begin{equation}
	L_{buffer}=\frac{T_{r}}{\frac{1}{R_{in}-R_{out}}+\frac{1}{R_{out}}}
	\label{eq3}
\end{equation}

The output rate can be easily determined, because the remote capture includes the arrival time for each packet and packet length is known. Input rate is not as trivial as output rate, because we only have the \textit{destination} capture. But if transmission time could be determined, for certain amount of packets, then input rate can also be found.

The times $ T_{r} $ and $ T_{t} $ are not the same because the buffer never empties completely, since new packets are always arriving at the input. However, $ t $ can be exactly measured and it will have the same value in both extremes. In this case, when \textit{destination} receives $ n $ consecutive packets, the \textit{source} has sent $ n+m $ packets (where $ m $ is the number of dropped packets). $ m $ can be known since all the packets have a unique identifier. With this information, output and input rate can be estimated only from the data contained in the \textit{destination} capture, using the following expression:

\begin{equation}
	R_{in}=\frac{n_{tx}+m_{tx}}{t} \times packet_{size} 
	\label{eq4}
\end{equation}

\section{Experimental results}

In order to illustrate the proposed methodology, real tests have been deployed in a testbed, which scheme is shown in figure \ref{fig:buffer_size_topology} and results are analysed based on procedures shown in figure \ref{fig:buffer_size} and \ref{fig:dropping}. Real machines have been used (Linux kernel $ 2.6.38-7 $, Atheros $ AR9287 $ wireless network adapter, $ Intel^{\circledR} \; Core^{TM} \; i3 \; CPU \; 2.4 \; GHz $) in a laboratory environment, in order to identify the buffer behaviour of different devices. 

\subsection{ETG (E2E Traffic Generator)}

This tool was developed by our group \cite{gtc13} with the aim of using it for link analysis in VoIP. In this work we will use it to generate the IP flows. It allows  us to calculate objective quality parameters (delay, jitter and loss rate) and also subjective ones (R factor, MOS). It is capable of establishing E2E communications between hosts, by sending and receiving multiple UDP traffic bursts. It generates one-way and round trip traffic and captures it at the ends of the communication. 

One advantage of this application is that it provides repeatability for the tests. In addition, a mechanism allows task automation, allowing test iteration between start and end moments. It is also able to generate traffic emulating the characteristics of certain services, and it allows sending packets whose size and transmission intervals are read from  a file.

\subsection{Real scenarios}

We have studied two different scenarios, in both cases a host sends traffic to a destination through a network or a device. A sniffer is included at the best location for making captures. ETG was used to generate UDP flows to flood the buffer, and also to automate \textit{tcpdump} captures in the ends of the link. Flows were sent from \textit{source} to \textit{destination}. Different bandwidths were set in the hubs in order to create a bottleneck which has to be measured. Flow control options in the switch are disabled.

\subsubsection{First scenario: Wireless network}

\begin{figure}
	\centering
	\includegraphics[width=3.5in]{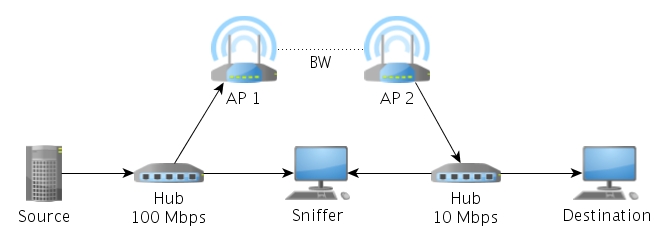}
	\caption{First scenario: Estimating buffer size in a wireless network.}
	\label{fig:topology_wifi}
\end{figure}

In the first case, a link between two access points (Linksys WAP54G) has been analized, trying to estimate the buffer size of this access point. Figure \ref{fig:topology_wifi} shows the topology used: both hosts are connected to the hubs using $ 100 \; Mbps $ and $ 10 \; Mbps $ links respectively. UDP flows were sent from \textit{source} to \textit{destination}, with the aim of obtaining the buffer size of $ AP 1 $.

\subsubsection{Second scenario: Wired network}

\begin{figure}
	\centering
	\includegraphics[width=3.5in]{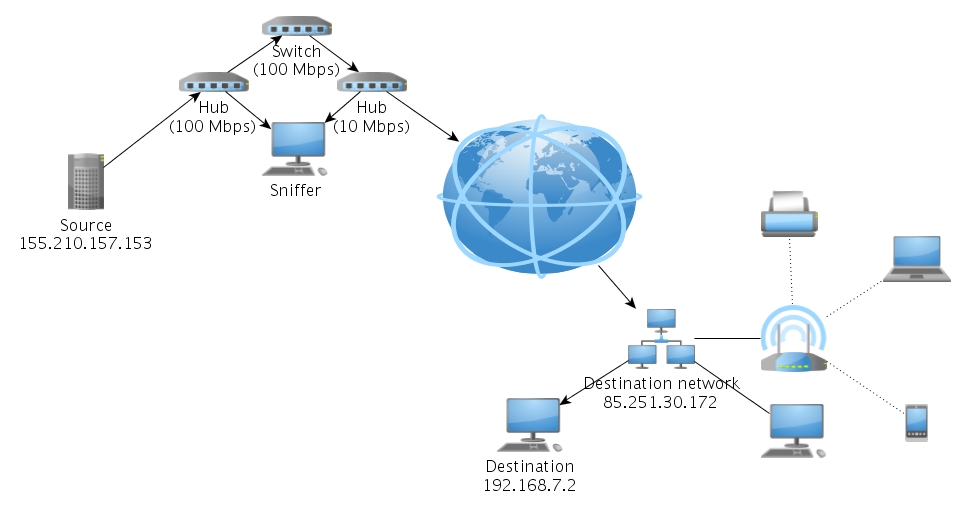}
	\caption{Second scenario: Estimating buffer size in a wired network.}
	\label{fig:topology_ethernet}
\end{figure}

For the second scenario, the topology shown in figure \ref{fig:topology_ethernet} has been used. In this case, a typical home network is accessed by other host from a different network across the internet. UDP flows were sent from the \textit{source} to the \textit{destination}. This scenario permits partial access to the SUT, so the measurements can estimate switch's buffer size by packets going from \textit{source} to \textit{destination}. Using the remote capture buffer size, the concatenation of different buffers across the internet can be estimated.

\subsection{Measurements with physical and remote access}

\begin{figure}
	\centering
	\includegraphics[width=3in]{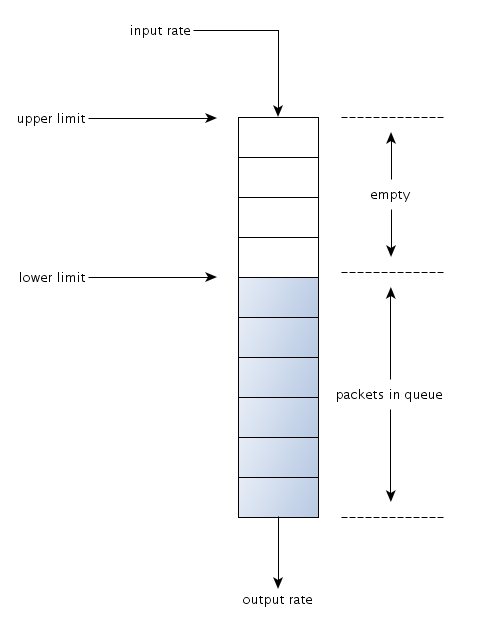}
	\caption{Buffer characteristics.}
	\label{fig:buffer}
\end{figure}

A particular buffer behaviour has been observed in the wired and wireless devises tested in both scenarios: when the buffer is completely full no more packets are accepted, it will be called \textit{upper limit}. In this moment, the buffer does not accept new packets until a certain amount of memory is available, it will be called \textit{lower limit}, and in this moment the filling process begins again. This can be seen in figure \ref{fig:buffer}. 

Although this behaviour has some similarities with Random Early Detection (RED), it is not the same: when RED is used, if the buffer is almost empty, all incoming packets are accepted but when the number of enqueued packets grows, the probability for dropping an incoming packet grows too. Finally, if the buffer is full, all incoming packets are dropped. On the other hand, the observed buffer behaviour does not use any dropping probability: if the buffer is full, it does not accept new packets until a certain amount of memory is available.

It is well known \cite{buffers1} that some manufacturers measure the buffer in number of packets whereas others define it in number of bytes, or even in terms of the maximum allowed queueing delay. In order to take this into account, UDP flows using three different packet lengths are sent to flood the buffer. If the estimating size is the same for the three traffics, then we know that the buffer is measured in number of packets. Otherwise, it is measured in bytes. Table \ref{table:interarrival_times} presents a summary of the traffic used for these tests.

\begin{table}
\renewcommand{\arraystretch}{1.3}

\caption{Inter-Packet time in $ \mu s $ }
\label{table:interarrival_times}
\centering
\scalebox{1}[1.2]{

	\begin{tabular}{ccccc}
		\hline
		\hline
		$ Packet size $ & $ 10\;Mbps $ & $ 20\;Mbps $ & $ 30\;Mbps $ & $ 40\;Mbps $\\
		\hline
		$ 1500 $ & $ 1200 $ & $ 600 $ & $ 400 $ & $ 300 $\\
		$ 800 $ & $ 640 $ & $ 320 $ & $ 213.3 $ & $ 160 $\\
		$ 200 $ & $ 160 $ & $ 80 $ & $ 53.3 $ & $ 40 $\\
		\hline
		\hline
	\end{tabular}
}
\end{table}

For both scenarios, buffer size is the same when changes in the packet length are made, so in these cases, buffer size is defined by number of packets.

\subsubsection{First scenario}

In this  wireless scenario we have compared the two methods when there is physical access to the SUT (table \ref{table:buffer_wifi}). The presented results are the ones obtained using packets of $ 1500 $ bytes, since they are the most accurate. It can be seen that \textit{method 1} is the most accurate estimation so it has been used to compare with \textit{method 2}. Variations of the output rate in wireless network generate error growth. The reason for this is illustrated in table \ref{table:emptying_rate}: while filling rate is relatively constant, the emptying rate shows variations for the highest bandwidths. The reason for this is that the WiFi access point switches from higher to lower speeds depending on the state of the radio channel.
 
\begin{table}
	\renewcommand{\arraystretch}{1.3}	
	\centering
	\caption{Wifi access point buffer size for different bandwidth.}
	\label{table:buffer_wifi}
	\scalebox{0.7}[1.2]{
	\begin{tabular}{ccccccccccc}
		\hline
		\hline
		\multicolumn{2}{c}{$ Access\; point \; bandwidth $} & &  \multicolumn{2}{c}{$ Method \; 1 $} & & \multicolumn{2}{c}{$ Method \; 2 $} & & \multicolumn{2}{c}{$ Method \; 2 \; error \; (\%) $} \\ 
		\multicolumn{2}{c}{$ Mbps $} & & $ LL $ & $ UL $ & & $ LL $ & $ UL $ & & $ LL $ & $ UL $ \\
		\hline
		\multirow{6}{2.2cm}{$ Wi-Fi $} & $ 1 $ & & $ 30 $ & $ 55 $ & & $ 30 $ & $ 55 $ & & $ 0 $ & $ 0 $ \\ 
		 & $ 2 $ & & $ 30 $ & $ 55 $ & & $ 30 $ & $ 55 $ & & $ 0 $ & $ 0 $ \\ 
		 & $ 5.5 $ & & $ 30 $ & $ 55 $ & & $ 30 $ & $ 55 $ & & $ 0 $ & $ 0 $ \\ 
		 & $ 11 $ & & $ 30 $ & $ 55 $ & & $ 32 $ & $ 53 $ & & $ 6.67 $ & $ 3.63 $ \\ 
		 & $ 24 $ & & $ 30 $ & $ 55 $ & & $ 33 $ & $ 52 $ & & $ 10 $ & $ 5.45 $ \\ 
		 & $ 54 $ & & $ 30 $ & $ 55 $ & & $ 36 $ & $ 59 $ & & $ 20 $ & $ 7.27 $ \\  
		\hline 
		\hline
		{\footnotesize LL: Stands for lower limit}.&&&&&&&&&&\\
		{\footnotesize UL: Stands for upper limit.}&&&&&&&&&&\\
	\end{tabular}
	}
\end{table}

\begin{table}
	\renewcommand{\arraystretch}{1.3}	
	\centering
	\caption{Observed variations in buffer output rate for different bandwidth.}
	\label{table:emptying_rate}
	\scalebox{0.7}[1.2]{
	\begin{tabular}{ccccccc}
		\hline
		\hline
		$ Access \; point \; rate $ & $ 54 \; Mbps $ & $ 24 \; Mbps $ & $ 11 \; Mbps $ & $ 5.5 \; Mbps $ & $ 2 \; Mbps $ & $ 1 \; Mbps $ \\ 
		\hline
		$ Minimum (Mbps) $ & $ 10.88 $ & $ 13.7 $ & $ 5.75 $ & $ 2.29 $ & $ 1.24 $ & $ 0.65 $ \\ 
		$ Maximum (Mbps) $ & $ 28.36 $ & $ 16.84 $ & $ 6 $ & $ 3.13 $ & $ 1.41 $ & $ 0.65 $ \\
		\hline 
		\hline
	\end{tabular} 
	}
\end{table}

\subsubsection{Second scenario}

In  this case we want to test the accuracy of our method when there is no possibility of physical access to the SUT. In order to be able to present comparative studies, first of all, we obtained the buffer size using \textit{method 1} with physical access. Next, the estimations obtained using \textit{method 2} are compared with previous results, and the relative error is presented. These tests were deployed using physical and remote access (see table \ref{table:buffer_ethernet}). Again, the best results are obtained using $ 1500 $ bytes packets, so we present them. Three different amounts of bandwidth have been used in order to flood the buffer.

If we look at the \textit{physical access} results (first three rows of the table), we see that the accuracy of the buffer size estimation using the \textit{method 2} is high. In addition, the error diminishes as input rate grows. 

Regarding \textit{remote access} results, it should be noticed that the results are less accurate but they are still acceptable. As integer values for the buffer capacity are used, in some cases the results are exactly the same than the ones of \textit{method 1}, so the obtained value of the error is null.

Again, the values presented in the tables are the ones obtained using packets of $ 1500 $ bytes. The results using smaller packets are less accurate so we have not presented them.

\begin{table}
	\renewcommand{\arraystretch}{1.3}	
	\centering
	\caption{Ethernet buffer size for different bandwidth.}
	\label{table:buffer_ethernet}
	\scalebox{0.7}[1.2]{
	\begin{tabular}{ccccccccccc}
		\hline
		\hline
		\multicolumn{2}{c}{$ Bandwidth $} & & \multicolumn{2}{c}{$ Method \; 1 $} & & \multicolumn{2}{c}{$ Method \; 2 $} & & \multicolumn{2}{c}{$ Method \; 2 \; error \; (\%) $} \\
		\multicolumn{2}{c}{$ Mbps $} & & $ LL $ & $ UL $ & & $ LL $ & $ UL $ & & $ LL $ & $ UL $ \\
		\hline 
		\multirow{3}{2.2cm}{$ Ethernet \; (physical) $} & $ 20 $ & & $ 85 $ & $ 115 $ & & $ 83 $ & $ 113 $ & & $ 2.35 $ & $ 1.74 $ \\ 
		 & $ 30 $ & & $ 85 $ & $ 115 $ & & $ 84 $ & $ 114 $ & & $ 1.17 $ & $ 0.87 $ \\ 
		 & $ 40 $ & & $ 85 $ & $ 115 $ & & $ 84 $ & $ 115 $ & & $ 1.17 $ & $ 0 $ \\ 
		\hline
		\multirow{3}{2.2cm}{$ Ethernet \; (remote) $} & $ 20 $ & & $ 85^{*} $ & $ 115^{*} $ & & $ 80 $ & $ 111 $ & & $ 5.89 $ & $ 3.48 $ \\ 
		 & $ 30 $ & & $ 85^{*} $ & $ 115^{*} $ & & $ 83 $ & $ 113 $ & & $ 2.35 $ & $ 1.74 $ \\ 
		 & $ 40 $ & & $ 85^{*} $ & $ 115^{*} $ & & $ 84 $ & $ 115 $ & & $ 1.17 $ & $ 0 $ \\
		\hline 
		\hline
		\multicolumn{4}{c}{$ LL: \; Stands \; for \; lower \; limit. $}&&&&&&&\\
		\multicolumn{4}{c}{$ UL: \; Stands \; for \; upper \; limit. $}&&&&&&&\\
		\multicolumn{10}{c}{$ ^{*} These \; values \; were \; calculate \; as \; physical \; access \; to \; compare \; method \; 2. $}&\\
	\end{tabular}
	}
\end{table}

\section{Conclusion}

This article has presented two methods to analyse the technical and functional characteristics of commercial buffers of different devices, or even networks. This characterization is important, taking into account that the buffer may modify traffic characteristics, and may also drop packets.

The methodology can be used if there is physical access to the ``System Under Test", but it is also useful, with certain limitations, for measuring a remote system. Tests using commercial devices have been deployed in two different scenarios, using wired and wireless networks. A particular buffer behaviour has been observed for a device: once the buffer is full, it does not accept new packets until a certain space is again available.

The results show that accurate results of the buffer size can be obtained when there is physical access to the ``System Under Test". In case of having no direct access to the system, an acceptable estimation can also be obtained if the input rate is more than three times the output rate. In this case, big packets have to be used for the tests. As future work the method has to be improved in order to minimize the error, especially when measuring wireless devices.


\section*{Acknowledgment}

This work has been partially financed by CPUFLIPI Project (MICINN TIN2010-17298), MBACToIP Project, of Aragon I+D Agency, Ibercaja Obra Social, NDCIPI-QQoE Project of C\'atedra Telef\'onica, Univ. of Zaragoza and Fundaci\'on Carolina.
\nocite{games1} \nocite{games2} \nocite{games3} \nocite{otros1} \nocite{otros4} \nocite{bandwidth1}

%
\bibliographystyle{IEEEtran}
\bibliography{citass}

\begin{thebibliography}{10}
\providecommand{\url}[1]{#1}
\csname url@samestyle\endcsname
\providecommand{\newblock}{\relax}
\providecommand{\bibinfo}[2]{#2}
\providecommand{\BIBentrySTDinterwordspacing}{\spaceskip=0pt\relax}
\providecommand{\BIBentryALTinterwordstretchfactor}{4}
\providecommand{\BIBentryALTinterwordspacing}{\spaceskip=\fontdimen2\font plus
\BIBentryALTinterwordstretchfactor\fontdimen3\font minus
  \fontdimen4\font\relax}
\providecommand{\BIBforeignlanguage}[2]{{%
\expandafter\ifx\csname l@#1\endcsname\relax
\typeout{** WARNING: IEEEtran.bst: No hyphenation pattern has been}%
\typeout{** loaded for the language `#1'. Using the pattern for}%
\typeout{** the default language instead.}%
\else
\language=\csname l@#1\endcsname
\fi
#2}}
\providecommand{\BIBdecl}{\relax}
\BIBdecl

\bibitem{buffers11}
\BIBentryALTinterwordspacing
R.~Stanojevi\'{c} and R.~Shorten, ``Trading link utilization for queueing
  delays: An adaptive approach,'' \emph{Comput. Commun.}, vol.~33, no.~9, pp.
  1108--1121, Jun. 2010. [Online]. Available:
  \url{http://dx.doi.org/10.1016/j.comcom.2010.02.014}
\BIBentrySTDinterwordspacing

\bibitem{buffers6}
\BIBentryALTinterwordspacing
C.~Villamizar and C.~Song, ``High performance tcp in ansnet,'' \emph{SIGCOMM
  Comput. Commun. Rev.}, vol.~24, pp. 45--60, October 1994. [Online].
  Available: \url{http://doi.acm.org/10.1145/205511.205520}
\BIBentrySTDinterwordspacing

\bibitem{buffers7}
\BIBentryALTinterwordspacing
G.~Appenzeller, I.~Keslassy, and N.~McKeown, ``Sizing router buffers,''
  \emph{SIGCOMM Comput. Commun. Rev.}, vol.~34, pp. 281--292, August 2004.
  [Online]. Available: \url{http://doi.acm.org/10.1145/1030194.1015499}
\BIBentrySTDinterwordspacing

\bibitem{buffers10}
\BIBentryALTinterwordspacing
M.~Enachescu, Y.~Ganjali, A.~Goel, N.~McKeown, and T.~Roughgarden, ``Part iii:
  routers with very small buffers,'' \emph{SIGCOMM Comput. Commun. Rev.},
  vol.~35, pp. 83--90, July 2005. [Online]. Available:
  \url{http://doi.acm.org/10.1145/1070873.1070886}
\BIBentrySTDinterwordspacing

\bibitem{buffers9}
\BIBentryALTinterwordspacing
A.~Vishwanath and V.~Sivaraman, ``{Routers With Very Small Buffers: Anomalous
  Loss Performance for Mixed Real-Time and TCP Traffic},'' pp. 80--89, Jun.
  2008. [Online]. Available: \url{http://dx.doi.org/10.1109/IWQOS.2008.16}
\BIBentrySTDinterwordspacing

\bibitem{buffers5}
\BIBentryALTinterwordspacing
A.~Vishwanath, V.~Sivaraman, and M.~Thottan, ``Perspectives on router buffer
  sizing: recent results and open problems,'' \emph{SIGCOMM Comput. Commun.
  Rev.}, vol.~39, pp. 34--39, March 2009. [Online]. Available:
  \url{http://doi.acm.org/10.1145/1517480.1517487}
\BIBentrySTDinterwordspacing

\bibitem{buffers1}
\BIBentryALTinterwordspacing
J.~Sommers, P.~Barford, A.~Greenberg, and W.~Willinger, ``An sla perspective on
  the router buffer sizing problem,'' \emph{SIGMETRICS Perform. Eval. Rev.},
  vol.~35, pp. 40--51, March 2008. [Online]. Available:
  \url{http://doi.acm.org/10.1145/1364644.1364645}
\BIBentrySTDinterwordspacing

\bibitem{buffers2}
\BIBentryALTinterwordspacing
A.~Vishwanath, V.~Sivaraman, and G.~N. Rouskas, ``Considerations for sizing
  buffers in optical packet switched networks,'' \emph{IEEE INFOCOM 2009 The
  28th Conference on Computer Communications}, pp. 1323--1331, 2009. [Online].
  Available:
  \url{http://ieeexplore.ieee.org/lpdocs/epic03/wrapper.htm?arnumber=5062047}
\BIBentrySTDinterwordspacing

\bibitem{buffers3}
\BIBentryALTinterwordspacing
A.~Lakshmikantha, R.~Srikant, and C.~Beck, ``Impact of file arrivals and
  departures on buffer sizing in core routers,'' \emph{IEEE INFOCOM The 27th
  Conference on Computer Communications}, vol.~19, pp. 86--90, 2008. [Online].
  Available:
  \url{http://ieeexplore.ieee.org/lpdocs/epic03/wrapper.htm?arnumber=4509621}
\BIBentrySTDinterwordspacing

\bibitem{buffers4}
\BIBentryALTinterwordspacing
A.~Dhamdhere and C.~Dovrolis, ``Open issues in router buffer sizing,''
  \emph{SIGCOMM Comput. Commun. Rev.}, vol.~36, pp. 87--92, January 2006.
  [Online]. Available: \url{http://doi.acm.org/10.1145/1111322.1111342}
\BIBentrySTDinterwordspacing

\bibitem{buffers8}
\BIBentryALTinterwordspacing
N.~Beheshti, Y.~Ganjali, M.~Ghobadi, N.~McKeown, and G.~Salmon, ``Experimental
  study of router buffer sizing,'' pp. 197--210, 2008. [Online]. Available:
  \url{http://doi.acm.org/10.1145/1452520.1452545}
\BIBentrySTDinterwordspacing

\bibitem{gtc14}
J.~Saldana, J.~Fern\'{a}ndez-Navajas, J.~Ruiz-Mas, E.~Viruete~Navarro, and
  L.~Casadesus, ``Influence of online games traffic multiplexing and router
  buffer on subjective quality,'' \emph{in Proc. CCNC 2012- 4th IEEE
  International Workshop on Digital Entertainment, Networked Virtual
  Environments, and Creative Technology (DENVECT)}, pp. 482--486, Las Vegas,
  January 2012.

\bibitem{gtc15}
J.~Saldana, J.~Murillo, J.~Fern\'{a}ndez-Navajas, J.~Ruiz-Mas, E.~Viruete, and
  J.~I. Aznar, ``Evaluation of multiplexing and buffer policies influence on
  voip conversation quality,'' \emph{In Proc. CCNC 2011- 3rd IEEE International
  Workshop on Digital Entertainment, Networked Virtual Environments, and
  Creative Technology}, pp. 1147--1151, Las Vegas, January 2011.

\bibitem{gtc13}
L.~A. Casadesus~Pazos, J.~Fern\'{a}ndez~Navajas, J.~Ruiz~Mas, J.~M.
  Saldana~Medina, J.~I. Aznar~Baranda, and E.~Viruete~Navarro, ``Herramienta
  para automatizaci\'{o}n de medidas de tiempo real extremo a extremo,''
  \emph{Actas del XXVI Simposium Nacional de la Uni\'{o}n Cient\'{i}fica
  Internacional de Radio (URSI 2011)}, Legan\'{e}s (Espa\~{n}a). ISBN
  9788493393458. Sept. 2011.

\bibitem{games1}
M.~Ries, P.~Svoboda, and M.~Rupp, ``Empirical study of subjective quality for
  massive multiplayer games,'' in \emph{Systems, Signals and Image Processing,
  2008. IWSSIP 2008. 15th International Conference on}, june 2008, pp. 181
  --184.

\bibitem{games2}
P.~Svoboda, W.~Karner, and M.~Rupp, ``Traffic analysis and modeling for world
  of warcraft,'' in \emph{Communications, 2007. ICC '07. IEEE International
  Conference on}, june 2007, pp. 1612 --1617.

\bibitem{games3}
G.~Huang, M.~Ye, and L.~Cheng, ``Modeling system performance in mmorpg,'' in
  \emph{Global Telecommunications Conference Workshops, 2004. GlobeCom
  Workshops 2004. IEEE}, nov.-3 dec. 2004, pp. 512 -- 518.

\bibitem{otros1}
\BIBentryALTinterwordspacing
A.~Golaup and H.~Aghvami, ``A multimedia traffic modeling framework for
  simulation-based performance evaluation studies,'' \emph{Computer Networks},
  vol.~50, no.~12, pp. 2071 -- 2087, 2006, network Modelling and Simulation.
  [Online]. Available:
  \url{http://www.sciencedirect.com/science/article/pii/S1389128605003294}
\BIBentrySTDinterwordspacing

\bibitem{otros4}
\BIBentryALTinterwordspacing
A.~Lombardo, G.~Morabito, and G.~Schembra, ``Statistical traffic modeling and
  guaranteed service disciplines: a performance evaluation paradigm,''
  \emph{Computer Networks}, vol.~36, no. 5-6, pp. 579 -- 595, 2001. [Online].
  Available:
  \url{http://www.sciencedirect.com/science/article/pii/S1389128601001700}
\BIBentrySTDinterwordspacing

\bibitem{bandwidth1}
\BIBentryALTinterwordspacing
K.~Lakshminarayanan, V.~N. Padmanabhan, and J.~Padhye, ``Bandwidth estimation
  in broadband access networks,'' in \emph{Proceedings of the 4th ACM SIGCOMM
  conference on Internet measurement}, ser. IMC '04.\hskip 1em plus 0.5em minus
  0.4em\relax New York, NY, USA: ACM, 2004, pp. 314--321. [Online]. Available:
  \url{http://doi.acm.org/10.1145/1028788.1028832}
\BIBentrySTDinterwordspacing

\end{thebibliography}


\end{document}